\documentclass[fleqn,usenatbib,letters]{mnras}

\usepackage{newtxtext,newtxmath}

\usepackage[T1]{fontenc}

\DeclareRobustCommand{\VAN}[3]{#2}
\let\VANthebibliography\thebibliography
\def\thebibliography{\DeclareRobustCommand{\VAN}[3]{##3}\VANthebibliography}

\usepackage{graphicx}	
\usepackage{amsmath}	
\usepackage{gensymb}
\usepackage{subcaption}
\title[A strongly lensed dual-AGN system]{PS\,J1721+8842: A gravitationally lensed dual AGN system at redshift 2.37 with two radio  components}

\author[C. S. Mangat et al.]{
C. S. Mangat,$^{1}$ 
J. P. McKean,$^{2,3}$\thanks{E-mail: mckean@astro.rug.nl} 
R. Brilenkov,$^{3}$ 
P. Hartley,$^{4}$ 
H. R. Stacey,$^{5}$ 
S. Vegetti$^{5}$ and
D. Wen$^{3}$
\\
$^1$Birla Institute of Technology and Science, Shamirpet-Keesara Road, Jawahar Nagar, Shameerpet, Hyderabad, Telangana 500078, India\\
$^2$ASTRON, Netherlands Institute for Radio Astronomy, Oude Hoogeveensedijk 4, 7991 PD, Dwingeloo, the Netherlands\\
$^3$Kapteyn Astronomical Institute, University of Groningen, P.O. Box 800, 9700AV Groningen, the Netherlands\\
$^4$SKAO, Jodrell Bank, Lower Withington, Macclesfield, Cheshire, SK11 9FT, UK\\
$^5$Max Planck Institute for Astrophysics, Karl-Schwarzschild Str. 1, D-85748 Garching bei München, Germany
}

\date{Accepted 2021 September 6. Received 2021 September 6; in original form 2021 August 10}

\pubyear{2021}

\begin{document}
\label{firstpage}
\pagerange{\pageref{firstpage}--\pageref{lastpage}}
\maketitle

\begin{abstract}
Dual-Active Galactic Nuclei (AGN) are a natural consequence of the hierarchical structure formation scenario, and can provide an important test of various models for black hole growth. However, due to their rarity and difficulty to find at high redshift, very few confirmed dual-AGN are known at the epoch where galaxy formation peaks. Here we report the discovery of a gravitationally lensed dual-AGN system at redshift 2.37 comprising two optical/IR quasars separated by $6.5\pm0.6$~kpc, and a third compact ($R_{\rm eff} = 0.45\pm0.02$~kpc) red galaxy that is offset from one of the quasars by $1.7\pm0.1$~kpc. From Very Large Array imaging at 3 GHz, we detect 600 and 340 pc-scale radio emission that is associated with both quasars. The 1.4 GHz luminosity densities of the radio sources are about $10^{24.35}$~W~Hz$^{-1}$, which is consistent with weak jets. However, the low brightness temperature of the emission is also consistent with star-formation at the level of 850 to 1150~M$_{\odot}$~yr$^{-1}$. Although this supports the scenario where the AGN and/or star-formation is being triggered through an ongoing triple-merger, a post-merger scenario where two black holes are recoiling is also possible, given that neither has a detected host galaxy.
\end{abstract}




\section{Introduction}
\label{sec:intro}

According to the hierarchical structure formation scenario, supermassive black holes (SMBHs) are created as a result of a major merger involving two massive galaxies, each with its own central black hole \citep{White_1978}. Furthermore, the major merger can push gas onto one or both of the black holes, providing a mechanism for triggering an active galactic nucleus (AGN) at the centre of one or both of the merging galaxies (\citealt{Di_Matteo_2005,Weston_2016,Ellison_2019}). However, recent studies, both from observations and simulations, have shown that major mergers may not play such a dominant role in SMBH growth and in triggering AGN activity \citep{Cisternas_2010,Sharma_2021, Silva_2021,Shah_2020}. Therefore, the formation and growth of SMBHs at the centres of massive galaxies is still unclear. 

Structures with two actively accreting AGN that are separated by between 100~pc and 10~kpc are generally called dual-AGN systems, detailed studies of which can be used to test the major merger scenario. This has led to an increased interest in these complex objects  through both new observations \citep{Koss_2012} and hydrodynamical simulations \citep{Rosas_Guevara_2018}. However, dual-AGN are extremely rare ($\sim 0.3$~percent of the AGN population; \citealt*{Volonteri_2003}), and their discovery has mostly occurred serendipitously at optical or X-ray wavelengths  (e.g. \citealt{Bentez_2013,Lena_2018}). Their rarity has been attributed to the short lifetime of simultaneous accretion in both of the black holes, and the inability to characterize the observed properties without extensive multi-wavelength follow up \citep{De_Rosa_2019}. Observations at radio wavelengths are particularly useful for studying dual-AGN, since these data are not affected by the large amounts of obscuring dust that are expected to be present in major mergers, and the high angular resolutions (milliarcsec scale) available can be used to identify the black holes \citep{Bondi_2010}. However, most AGN do not emit strongly at radio wavelengths, and so, dual-AGN where both components are radio-loud are rarer \citep{Rubinur_2019}.

Here, we present high resolution imaging at optical/infrared (IR) and radio wavelengths of the gravitational lens system PS\,J1721+8842, which was originally identified as a quadruply- or possibly a quintuply-imaged quasar at redshift 2.37 in the first and second data releases of {\it Gaia} \citep{Lemon_2018, Rusu_2019}. From these data, we find that PS\,J1721+8842 is comprised of two distinct quasars that are formed into two and four images, respectively. Both quasars are optically and radio luminous, and separated by 6.5~kpc in the source plane, making this object the first dual-AGN system that is also gravitationally lensed. Our letter is arranged as follows. In Sections~\ref{sec:observations} and \ref{sec:lens}, we present the observations and lens modelling, respectively. In Section~\ref{sec:discussion}, we discuss the various scenarios that can explain the complex morphology of PS\,J1721+8842, and outline future work to be carried out on this intriguing system.

Throughout, we assume a flat $\Lambda$CDM Universe with $H_0 = \rm{67.8~km\,s^{-1}~Mpc^{-1}}$, $\Omega_{\mathrm{M}} = 0.31$ and $\Omega_{\Lambda} = 0.69$ \citep{planck2016}. At the redshift of the two quasars, 1 arcsec corresponds to 8.53 kpc. We adopt a spectroscopic redshift of 0.184 (unpublished) for the foreground lensing galaxy.

\section{Observations}
\label{sec:observations}

In this section, we present {\it Hubble Space Telescope} ({\it HST}) and Karl G. Jansky Very Large Array (VLA) observations of PS\,J1721+8842 at optical/IR and radio wavelengths, respectively.

\begin{figure}
    \includegraphics[height=7.8cm]{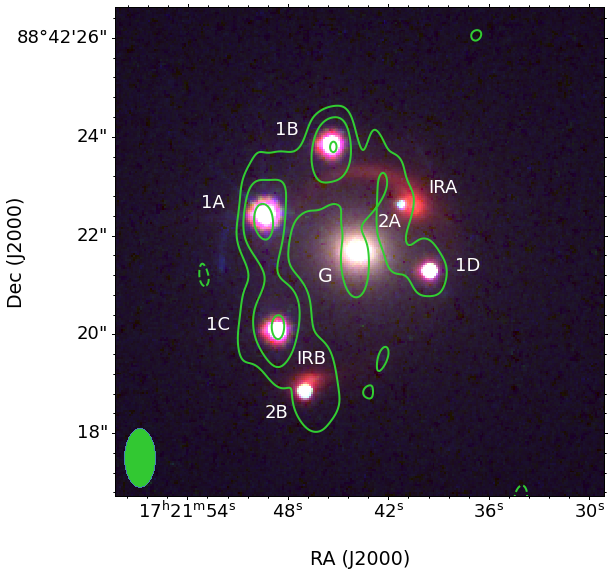}
    \caption{A pseudo-colour image of PS\,J1721+8842 obtained with the {\it HST}/WFC3 through the F475X, F814W and F160W filters. Overlaid (in green) are the 3 GHz VLA surface brightness contours. These data show that the system is comprised of a single massive elliptical galaxy at the centre, and has four lensed images (1A, 1B, 1C and 1D; Source 1) of one quasar and two lensed images (2A and 2B; Source 2) of a second quasar. Note that our labelling differs from that of \citet{Lemon_2018}, and is ordered by lensing magnification, where image A has the highest magnification. Also, there is a doubly-imaged IR bright galaxy (IRA and IRB) that is offset from the doubly imaged quasar. The quasar images of Source 1 have a red halo due to the different point spread function of the UVIS2 and IR cameras. The foreground lensing galaxy and the two lensed quasars also show evidence of radio emission. The 3 GHz surface brightness contours are at $(-3, 3, 6, 12) \times 6.2~\mu$Jy~beam$^{-1}$, the rms map noise. The VLA beam is shown in the lower left corner and has a size of $1.17 \times 0.64$ arcsec at a position angle of $-1$ deg East of North.}
    \label{fig:data_img}
\end{figure}

\subsection{\textit{Hubble Space Telescope}}
\label{sec:optical}

PS\,J1721+8842 was observed with the Wide Field Camera 3 (WFC3) using the UVIS2 and IR cameras on 2019 June 9 (GO:15652; PI: Treu). These data were taken through the F475X, F814W and F160W filters, with total integration times of 1382, 1428 and 2197 s, respectively. The calibrated flat-fielded images were obtained from the {\it HST} archive and processed in the standard way using the {\sc astrodrizzle} package within {\sc drizzlepac}. A pseudo-colour image of the gravitational lens system is presented in Fig.~\ref{fig:data_img}. 

We find that PS\,J1721+8842 has a complex surface brightness distribution at optical/IR wavelengths. At the centre of the system is the lensing galaxy (G), but in addition to the five lensed images that were found previously with {\it Gaia}, the new high resolution {\it HST} imaging detects a sixth quasar image. As we will demonstrate in the next section, these six images must be produced from two distinct quasars at the same redshift, where Source 1 results in lensed images 1A, 1B, 1C and 1D, and Source 2 results in lensed images 2A and 2B. We also detect an extended (observed-frame) IR component (IRA and IRB) close to, but offset to the west and north from lensed images 2A and 2B, respectively. We do not see any evidence of an extended gravitational arc or Einstein ring that is associated with the four lensed images of Source 1. 

To parameterize the optical emission (F475X and F814W) from PS\,J1721+8842, we fitted the quasar images and the lensing galaxy with the point spread function and a S\'ersic profile, respectively, using {\sc galfit} \citep{Peng_2010}. The relative positions and flux ratios of the lensed images from this analysis are presented in Table~\ref{tab:data_table}. As the IR data (F160W) has an extended lensed component, we instead parameterized this during the lens modelling (see Section~\ref{sec:lens}).

\subsection{Karl G.~Jansky Very Large Array}

We observed PS\,J1721+8842 with the VLA in A-configuration on 2019 October 3 as part of a programme to study the radio emission from quadruply-imaged gravitationally lensed quasars (19A-336; PI: McKean). The data were taken at a central frequency of 3~GHz through $16\times128$~MHz spectral windows, each with 64 spectral channels, and through both circular polarizations. The total on-source integration time was 8.5~min, with a visibility averaging time of 3~s. 3C286 was used as the absolute flux density, delay and bandpass calibrator, and J1639+8631 was used for determining the antenna based complex gains. The data were calibrated and imaged using the Common Astronomy Software Application ({\sc casa}) package. 

Given the sensitivity of the data, and the time and frequency sampling used, a wide-field image that included all sources within the VLA primary beam was first made. This was used to self-calibrate the target field to remove residual complex gain errors, after which, all of the field sources (except the target) were modelled and subtracted from the visibility dataset. PS\,J1721+8842 was then imaged using natural weighting. A contour map of the 3 GHz surface brightness distribution of the target is also shown in Fig.~\ref{fig:data_img}. 

We find that there is radio emission associated with all six lensed images, and also with the lensing galaxy. To parameterize the emission, we fitted 2 dimensional elliptical Gaussian functions to the various components. The results of this analysis are also presented in Table~\ref{tab:data_table}. In the case of Source 1, the 3 GHz surface brightness is highest at the location of the quasar images 1A, 1B and 1C, and is marginally resolved with deconvolved sizes of 0.8 to 1.2 arcsec. The lensed radio emission associated with Source 2 is resolved, suggesting that these components are extended, but is also offset by about 0.14 and 0.5 arcsec ($1.7\sigma$-level) from the corresponding quasar images. The total radio emission from the system is $S_{\rm 3~GHz} = 638 \pm 33~\mu$Jy. 

\begin{table*}
\caption{The relative positions and flux ratios of the lensed images, obtained from fitting point sources to the F475X and F814W optical imaging. Also presented are the properties of the (deconvolved) 2 dimensional elliptical Gaussian functions fitted to the 3 GHz radio imaging (in the image-plane). For image 1D, the model reduces to a delta function, so no size for the major and minor axes, or the position angle are reported. For comparison, we also give the predicted flux ratios from the lens model.} 
\centering
    \begin{tabular}{llllllll}
        \hline
         Band & Image & 1A & 1B & 1C & 1D & 2A & 2B \\
        \hline
        F475X &
         Relative RA (arcsec) & $\equiv 0.000$ & $+1.330\pm0.001$ & $+0.237\pm0.001$ & $+3.319\pm0.002$ & $+2.752\pm0.005$ & $+0.795\pm0.002$ \\
         &
         Relative Dec (arcsec) & $\equiv 0.000$ & $+1.401\pm0.001$ & $-2.347\pm0.001$ & $-1.144\pm0.002$ & $+0.198\pm0.005$ & $-3.581\pm0.002$ \\
         &
         Flux ratio & $\equiv 1.000$ & $0.463\pm0.009$ & $0.444\pm0.008$ & $0.201\pm0.005$ & $0.016\pm0.001$ & $0.103\pm0.004$ \\
        \hline
        F814W &
         Relative RA (arcsec) & $\equiv 0.000$ & $+1.331\pm0.001$ & $+0.236\pm0.001$ & $+3.320\pm0.002$ & $+2.750\pm0.003$ & $+0.796\pm0.002$ \\
         &
         Relative Dec (arcsec) & $\equiv 0.000$ & $+1.402\pm0.001$ & $-2.347\pm0.001$ & $-1.143\pm0.002$ & $+0.199\pm0.004$ & $-3.581\pm0.002$ \\
         &
         Flux ratio & $\equiv 1.000$ & $0.534\pm0.010$ & $0.516\pm0.009$ & $0.212\pm0.005$ & $0.024\pm0.002$ & $0.122\pm0.004$ \\
        \hline
        3 GHz &
         Relative RA (arcsec)               & $\equiv 0.00$ & $+1.30\pm0.04$    & $+0.24\pm0.03$    & $+3.41\pm0.05$    & $+2.62\pm0.07$    & $+1.34\pm0.24$ \\
         &
         Relative Dec (arcsec)              & $\equiv 0.00$ & $+1.33\pm0.10$    & $-2.15\pm0.09$    & $-1.08\pm0.16$    & $+0.15\pm0.29$    & $-3.46\pm0.17$ \\
         &
         Flux density ($\mu$Jy)             & $145\pm16$    & $129\pm18$        & $139\pm17$        & $25\pm10$         & $76\pm21$         & $107\pm31$ \\
         &
         Brightness ($\mu$Jy~beam$^{-1}$)   & $84\pm6$      & $68\pm7$          & $79\pm7$          & $26\pm6$          & $28\pm6$          & $25\pm6$ \\
         &
         Major FWHM (arcsec)                & $1.2\pm0.2$   & $1.0\pm0.3$       & $0.8\pm0.3$       & $\equiv 0$        & $2.2\pm0.8$       & $2.2\pm0.7$ \\
         &
         Minor FWHM (arcsec)                & $0.4\pm0.1$   & $0.6\pm0.4$       & $0.7\pm0.3$       & $\equiv 0$        & $0.5\pm0.4$       & $0.8\pm0.5$ \\
         &
         PA (degr.)                         & $176\pm6$     & $146\pm35$        & $150\pm78$        & $\equiv 0$        & $11\pm11$         & $110\pm28$ \\
         &
         Flux ratio                         & $\equiv 1.00$ & $0.89\pm0.16$     & $0.96\pm0.16$     & $0.17\pm0.07$     & $0.52\pm0.16$     & $0.74\pm0.23$ \\
         &
         Brightness ratio                   & $\equiv 1.00$ & $0.81\pm0.10$     & $0.94\pm0.11$     & $0.31\pm0.08$     & $0.33\pm0.08$     & $0.30\pm0.08$ \\
        \hline
        Lens model &
        Flux ratio                          & $\equiv 1.00$ & $0.86$     & $0.72$     & $0.28$     & $0.52$     & $0.36$ \\ 
        \hline
    \end{tabular}
	\label{tab:data_table}
\end{table*}

\section{Lens \& source modelling}
\label{sec:lens}

In this section, we present an overview of our modelling of the gravitational lensing data at optical/IR and radio wavelengths. This is needed to infer the intrinsic source properties of the lensed AGN. A more detailed description of the modelling of PS\,J1721+8842, and the other ten gravitational lens systems that comprise our VLA programme, will be presented by Brilenkov et al.~(in prep.).

\subsection{Optical/IR lens and source modelling}

\begin{figure}
    \centering
    \includegraphics[width=0.47\textwidth]{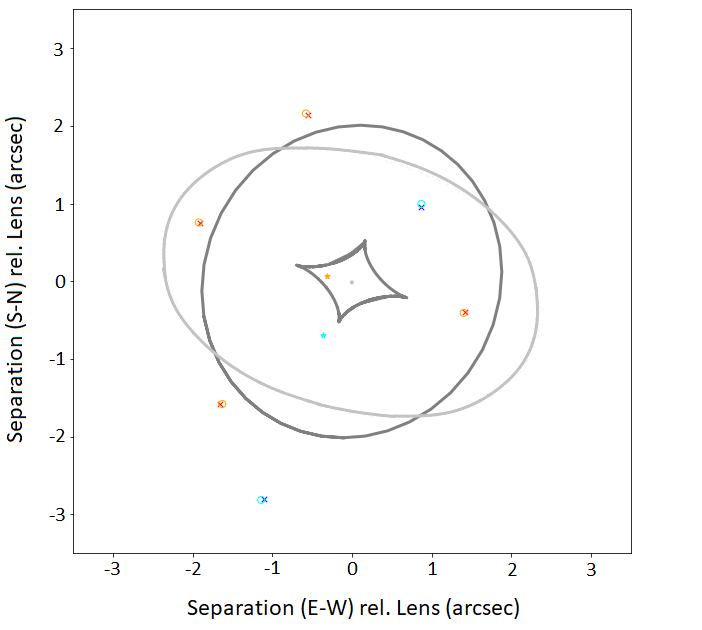}
    \caption{The best-fit SIE+shear lens model using the positions of all six lensed images (obtained using {\sc glafic}). The source-plane caustics are shown in dark grey, while the image-plane critical curves are shown in light grey. The crosses denote the observed image positions, while the circles denote the corresponding predicted image positions from this model. The source positions of the quasars are plotted as orange (Source 1) and cyan (Source 2) stars. The lens mass centre is at the origin.}
    \hspace{\fill}
    \label{fig:point_img}
\end{figure}

We modelled the gravitational lensing in two steps. First, using the observed image positions of the F475X and F814W data (see Table~\ref{tab:data_table}), we fitted a singular isothermal ellipsoid (SIE) plus an external shear mass distribution using {\sc glafic} \citep{Oguri_2010}. The flux-ratios of the quasar images were not used, as these are typically affected by intrinsic variability and/or microlensing \citep{Millon_2020}. However, the six image positions provide twelve observational constraints to the model, which has seven free parameters (the Einstein radius $\theta_E$; the lensing galaxy position $\Delta$RA, $\Delta$Dec; the ellipticity $e$ and the position angle PA; and the external shear $\Gamma$ and position angle $\Gamma_\theta$). The resulting model is shown in Fig.~\ref{fig:point_img}, and is a good fit to the data. 

Second, using the parameters of the initial mass model from {\sc glafic}, we then used {\sc lenstronomy} \citep{Birrer_2015, Birrer_2018} to model the gravitational lens mass and source surface brightness distribution using the F475X, F814W and F160W imaging data simultaneously. For this, we parameterized the light from the lens and the extended (IR) source component with S\'ersic profiles. The quasar images were modelled using the point spread functions of the UVIS2 and IR cameras to determine the positions, but similar to above, the flux of each image was not used as a constraint in the lens modelling and was left as a free parameter. The lens and source positions were first determined using Particle Swarm Optimization based ray-tracing, followed by a Monte-Carlo Markov Chain based optimization. This ensured an effective search of the parameter space and provided uncertainties on the fitted parameters for the given model. The mass model parameters for the lens, and the point- and extended-source models for the two quasars and the IR source component, respectively, are presented in Table~\ref{tab:param_table}. By considering the residual flux in the three {\it HST} images, we find that this model has a reduced-$\chi^2$ of 2.11, which is dominated by the residual light from the point sources. In Table~\ref{tab:image_table}, we present the predicted lensed image positions, magnifications and time-delays. We find that this model can recover the observed positions of the lensed images to within the measurement and modelling uncertainties.

\subsection{Radio source modelling}

Using the lens model obtained from the optical/IR data, we fitted the radio data, keeping all of the lens parameters fixed except for the position of the mass distribution (so that any mis-match between the {\it HST} and VLA absolute astrometry could be accounted for). The modelling was done in the visibilty plane using {\sc visilens} \citep{Spilker_2016}, and given the quality of the data, a circular Gaussian model was used to parameterize each source. We find that Source 1 and 2 have an effective radius of $R_{\rm eff}=0.07\pm0.01$ and $0.04\pm0.01$~arcsec, respectively, and a total magnification of $\mu_{\rm radio}=16.3$ and 5.1, respectively. However, we take these results only as indicative of the true source sizes, since further high resolution observations will be needed.

\subsection{Evidence for a flux-ratio anomaly}

From Table~\ref{tab:data_table}, we find that the observed optical flux-ratios change as a function of wavelength, and are not consistent with the expected values from the lens model. In particular, the flux-ratio for Source 2 is inverted. This is likely not an issue with the mass model, but suggests that the optical data are effected by microlensing. The radio data are not expected to be significantly affected by microlensing because the derived source sizes are too large. However, we find that image 1C and 1D show evidence for a flux-ratio anomaly at the $1.6\sigma$ level. If genuine, this could be due to intrinsic variability, a complex source structure, or due to a perturbation in the mass model. Further high resolution radio imaging is needed to investigate this.

\subsection{Source-plane properties}

\begin{figure}
    \centering
    \includegraphics[width=0.47\textwidth]{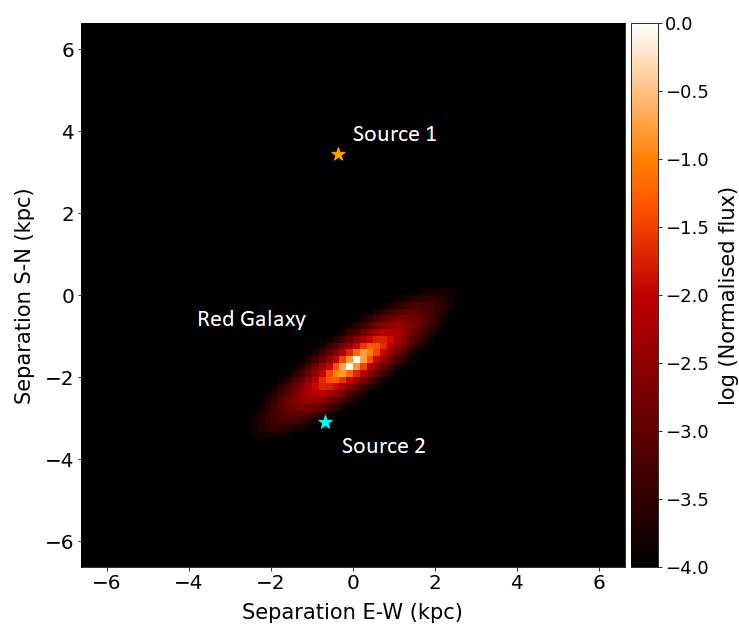}
    \caption{The reconstructed source-plane model showing the projected positions of Source 1, Source 2 and the IR component. This confirms that all three source components are at redshift 2.37, and are spatially offset. Note that the strong ellipticity of the IR component is not real, but is an artefact of the preferential magnification provided by the lens.}
    \hspace{\fill}
    \label{fig:source_img}
\end{figure}

In Fig.~\ref{fig:source_img}, we present a representation of the source-plane structure of PS\,J1721+8842 at redshift 2.37. From our lens modelling analysis, we confirm that all three optical/IR components are at the same redshift and offset from each other. The projected separation of the two quasars is $6.5\pm0.6$ kpc, confirming that this is a dual-AGN system. The IR component is offset from Source 2 by $1.6\pm0.3$ kpc in projection, consistent with being a third distinct object. The effective radius of the IR component is $1.67\pm0.08$~kpc. However, this is likely an upper-limit to the size as most of the magnification is in the tangential direction, as a result of which, this component appears highly elliptical in the source-plane. Using instead the minor axis of the elliptical profile gives an effective radius of $0.45\pm0.02$~kpc. This, coupled with the derived S\'ersic index of $n = 2$, suggests that this is a compact red galaxy. It is not clear whether the red colour is due to dust, or an evolved stellar population. Such red and compact host galaxies have been found for other lensed and non-lensed AGN \citep{Kocevski_2017,Spingola_2020}, but in these cases, the AGN is embedded at the centre of the host galaxy.

The radio emission from PS\,J1721+8842 can be due to weak jets associated with AGN activity or from star formation. Given that the radio emission has an inferred size of $600\pm85$ and $340\pm85$~pc for Source 1 and 2, respectively, and the coincidence with the quasars, it is likely that in both cases the emission is due to weak jets. We calculate the intrinsic source properties following \citet{McKean_2011}. The estimated rest-frame 1.4 GHz luminosity density (unlensed), assuming a radio spectral index of $\alpha = -0.7$, is $L_{\rm 1.4~GHz} = 10^{24.31\pm0.03}$ and $10^{24.43\pm0.08}$~W~Hz$^{-1}$, respectively for Source 1 and 2, which is consistent with weak radio jets. However, given the estimated source sizes, we find that the brightness temperatures have a limit of around $>10^3$~K, which would also be consistent with star formation. Under the assumption that all of the radio emission is due to star formation gives a star-formation rate of around 850 and 1150~M$_{\odot}$~yr$^{-1}$ for Source 1 and 2, respectively. This is significantly higher than is typically found for dusty star-forming galaxies and quasar host galaxies at this epoch \citep{Stacey_2018}. Also, given the compact sizes, the inferred star-formation rate intensities for the two quasars would be of order 800 and 3150~M$_{\odot}$~yr$^{-1}$~kpc$^{-2}$, which would imply greater than Eddington limited star formation. Further high resolution radio imaging and mm/spectral line imaging will be needed to confirm such an extreme starburst associated with the system.

\begin{table}
	\caption{Best fit SIE+shear mass model parameters (a), and the extended (b) and point source (c) models, obtained using {\sc lenstronomy}. All positions are relative to the centre of lens, and all position angles (PA) are given in degrees East of North.}
	\begingroup
	\renewcommand{\arraystretch}{1.06}
	\begin{subtable}[tb]{0.23\textwidth}
	\centering
	\begin{tabular}[tb]{ll}
		\hline
		Parameter & Best Fit\\
		\hline
		$\theta_E$ (arcsec)     & $1.9614\pm0.0002$\\
		$e$                     & $0.23\pm0.06$\\
		PA (degr.)              & $74\pm2$\\
		\hline
		$\Gamma$                & $0.07\pm0.01$\\
		$\Gamma_\theta$ (degr.) & $72\pm4$\\
		\hline
	\end{tabular}
	\subcaption{Lens mass model.}
	\label{tab:lens_table}
	\end{subtable}
	\endgroup
	\hspace{\fill}
	\begin{subtable}[tb]{0.24\textwidth}
	\centering
	\begin{tabular}[tb]{ll}
		\hline
		Parameter & Best fit\\
		\hline
		$\Delta$RA (arcsec)     & $-0.275\pm0.002$\\
		$\Delta$Dec (arcsec)    & $-0.543\pm0.003$\\
		$R_{\rm eff}$ (arcsec)  & $0.20\pm0.01$\\
		$n$                     & $2.03\pm0.05$\\
		$e$                     & $0.73\pm0.08$\\
		PA (degr.)      & $125\pm8$\\
		\hline
	\end{tabular}
	\subcaption{Extended source model.}
	\label{tab:host_table}
	\end{subtable}
	\begin{subtable}[h]{0.47\textwidth}
	\centering
	\begin{tabular}{lll}
		\hline
		Source & $\Delta$RA (arcsec) & $\Delta$Dec (arcsec)\\
		\hline
		Source 1 & $-0.32\pm0.03$ &  $+0.07\pm0.01$\\
		Source 2 & $-0.36\pm0.05$ & $-0.72\pm0.05$\\
		\hline
	\end{tabular}
	\subcaption{Point source model.}
	\label{tab:point_table}
	\end{subtable}
	\label{tab:param_table}
\end{table}

\begin{table}
	\centering
	\caption{The predicted image positions, magnifications ($\mu$; the sign donates the parity) and time-delays ($t_d$) for Source 1 and Source 2, obtained using {\sc lenstronomy}. All positions are relative to the centre of the lens.}
	\begin{tabular}{lcccr} 
		\hline
		Image & $\Delta$RA (arcsec) & $\Delta$Dec (arcsec) & $\mu$ & $t_d$ (days)\\
		\hline
		1A & $-1.8996\pm0.0004$ & $+0.7548\pm0.0008$ & $-5.8$ &0.0 \\
		1B & $-0.5694\pm0.0003$ & $+2.1538\pm0.0001$ & $+5.0$ &$-3.1$\\
		1C & $-1.6630\pm0.0002$ & $-1.5935\pm0.0002$ & $+4.2$ &$-5.9$\\
		1D & $+1.4193\pm0.0006$ & $-0.3905\pm0.0011$ & $-1.6$ &$+28.3$\\
		\hline
		2A & $+0.8501\pm0.0021$ & $+0.9573\pm0.0015$ & $-3.0$ &$+42.7$\\
		2B & $-1.1047\pm0.0005$ & $-2.8281\pm0.0003$ & $+2.1$ &$-45.4$\\
		\hline
	\end{tabular}
	\label{tab:image_table}
\end{table}

\section{Discussion \& Conclusions}
\label{sec:discussion}

We have shown that PS\,J1721+8842 is comprised of two quasars that form a dual-AGN system, with a third stellar component that forms part of a triple-merger. There is clear evidence of AGN activity in the form of two quasars and possibly a very high level of concentrated star-formation or weak radio jets within the quasar hosts. There are two scenarios that can explain the observed properties of PS\,J1721+8842.

First is a pre-merger scenario, wherein the active black holes (and their unseen host galaxies) are in the process of merging with the compact red galaxy. If this is the case, the radio emission at the location of the quasars could be dominated by AGN activity that has been possibly triggered through the interactions between the three structures. From studies of clusters and groups, it has been found that radio luminous AGN are more likely to be seen in less dense environments, and so galaxy interactions may play a more important role in their evolution, as opposed to direct mergers (e.g.~\citealt{Shen_2021}). However, this would go against the predictions from simulations that place dual-AGN activity in the latter phases of a merger \citep{Solanes_2019}. Also, we note that close-pair galaxy interactions can increase the (specific) star-formation rates of galaxies by a factor of a few \citep{Scott_2014}. Although we have no estimate of the stellar mass of the quasar hosts, the star-formation rates implied from the radio emission are likely too large to be consistent with a pre-merger phase.

Alternatively, a large star-formation rate with dual-AGN activity would be consistent with a post-merger scenario \citep{Sanders_1988}. However, there is no evidence of a host galaxy associated with either quasar, which is unusual as the highest magnification quasar images typically have an underlying extended stellar component \citep{Shajib_2019}. This leads to a somewhat speculative post-merger scenario, where the black holes have been `kicked out' of the host galaxy \citep{Volonteri_2020}. There have been several recent studies of SMBH recoil within quasars \citep{Shields_2008,Bonning_2007}, in mergers and AGN \citep{Blecha_2011}, and on their possible detection \citep{Raffai_2015,Blecha_2016}. Such a recoil would explain the offset in the position of Source 2 with respect to the centre of the IR component, which would be the host galaxy of the system. We estimate that the size of the host galaxy of Source 1 would have to be $<400$~pc to remain undetected in the {\it HST} imaging. If there continues to be no evidence of a host galaxy in further studies, then the SMBH recoil scenario could be invoked to explain why an active black hole is a few kpc outside its host galaxy. 

In summary, PS\,J1721+8842 will provide an important test of models for SMBH formation, and possibly the triggering of star formation. Further high-angular resolution observations at cm/mm wavelengths will be needed to understand the nature of the three components of the system, the origin of their radio emission, determine how they are dynamically linked, and establish whether this is a pre- or post-merger system.

\section*{Acknowledgements}
JPM acknowledges support from the Netherlands Organization for Scientific Research (NWO) (Project No. 629.001.023) and the Chinese Academy of Sciences (CAS) (Project No. 114A11KYSB20170054). This research is based on observations made with the NASA/ESA Hubble Space Telescope obtained from the Space Telescope Science Institute, which is operated by the Association of Universities for Research in Astronomy, Inc., under NASA contract NAS 5–26555. These observations are associated with program 15652. The National Radio Astronomy Observatory is a facility of the National Science Foundation operated under cooperative agreement by Associated Universities, Inc.

\section*{Data Availability}
The raw data used for our analysis is publicly available through the {\it HST} and VLA archives. 



\bibliographystyle{mnras}
\bibliography{psj1721} 

\begin{thebibliography}{}
\makeatletter
\relax
\def\mn@urlcharsother{\let\do\@makeother \do\$\do\&\do\#\do\^\do\_\do\%\do\~}
\def\mn@doi{\begingroup\mn@urlcharsother \@ifnextchar [ {\mn@doi@}
  {\mn@doi@[]}}
\def\mn@doi@[#1]#2{\def\@tempa{#1}\ifx\@tempa\@empty \href
  {http://dx.doi.org/#2} {doi:#2}\else \href {http://dx.doi.org/#2} {#1}\fi
  \endgroup}
\def\mn@eprint#1#2{\mn@eprint@#1:#2::\@nil}
\def\mn@eprint@arXiv#1{\href {http://arxiv.org/abs/#1} {{\tt arXiv:#1}}}
\def\mn@eprint@dblp#1{\href {http://dblp.uni-trier.de/rec/bibtex/#1.xml}
  {dblp:#1}}
\def\mn@eprint@#1:#2:#3:#4\@nil{\def\@tempa {#1}\def\@tempb {#2}\def\@tempc
  {#3}\ifx \@tempc \@empty \let \@tempc \@tempb \let \@tempb \@tempa \fi \ifx
  \@tempb \@empty \def\@tempb {arXiv}\fi \@ifundefined
  {mn@eprint@\@tempb}{\@tempb:\@tempc}{\expandafter \expandafter \csname
  mn@eprint@\@tempb\endcsname \expandafter{\@tempc}}}

\bibitem[\protect\citeauthoryear{{Ben{\'\i}tez} et~al.,}{{Ben{\'\i}tez}
  et~al.}{2013}]{Bentez_2013}
{Ben{\'\i}tez} E.,  et~al., 2013, \mn@doi [\apj] {10.1088/0004-637X/763/1/36},
  \href {https://ui.adsabs.harvard.edu/abs/2013ApJ...763...36B} {763, 36}

\bibitem[\protect\citeauthoryear{{Birrer} \& {Amara}}{{Birrer} \&
  {Amara}}{2018}]{Birrer_2018}
{Birrer} S.,  {Amara} A.,  2018, \mn@doi [Physics of the Dark Universe]
  {10.1016/j.dark.2018.11.002}, \href
  {https://ui.adsabs.harvard.edu/abs/2018PDU....22..189B} {22, 189}

\bibitem[\protect\citeauthoryear{{Birrer}, {Amara}  \& {Refregier}}{{Birrer}
  et~al.}{2015}]{Birrer_2015}
{Birrer} S.,  {Amara} A.,   {Refregier} A.,  2015, \mn@doi [\apj]
  {10.1088/0004-637X/813/2/102}, \href
  {https://ui.adsabs.harvard.edu/abs/2015ApJ...813..102B} {813, 102}

\bibitem[\protect\citeauthoryear{{Blecha}, {Cox}, {Loeb}  \&
  {Hernquist}}{{Blecha} et~al.}{2011}]{Blecha_2011}
{Blecha} L.,  {Cox} T.~J.,  {Loeb} A.,   {Hernquist} L.,  2011, \mn@doi
  [\mnras] {10.1111/j.1365-2966.2010.18042.x}, \href
  {https://ui.adsabs.harvard.edu/abs/2011MNRAS.412.2154B} {412, 2154}

\bibitem[\protect\citeauthoryear{{Blecha} et~al.,}{{Blecha}
  et~al.}{2016}]{Blecha_2016}
{Blecha} L.,  et~al., 2016, \mn@doi [\mnras] {10.1093/mnras/stv2646}, \href
  {https://ui.adsabs.harvard.edu/abs/2016MNRAS.456..961B} {456, 961}

\bibitem[\protect\citeauthoryear{{Bondi} \& {P{\'e}rez-Torres}}{{Bondi} \&
  {P{\'e}rez-Torres}}{2010}]{Bondi_2010}
{Bondi} M.,  {P{\'e}rez-Torres} M.~A.,  2010, \mn@doi [\apjl]
  {10.1088/2041-8205/714/2/L271}, \href
  {https://ui.adsabs.harvard.edu/abs/2010ApJ...714L.271B} {714, L271}

\bibitem[\protect\citeauthoryear{{Bonning}, {Shields}  \&
  {Salviander}}{{Bonning} et~al.}{2007}]{Bonning_2007}
{Bonning} E.~W.,  {Shields} G.~A.,   {Salviander} S.,  2007, \mn@doi [\apjl]
  {10.1086/521674}, \href
  {https://ui.adsabs.harvard.edu/abs/2007ApJ...666L..13B} {666, L13}

\bibitem[\protect\citeauthoryear{{Cisternas} et~al.,}{{Cisternas}
  et~al.}{2011}]{Cisternas_2010}
{Cisternas} M.,  et~al., 2011, \mn@doi [\apj] {10.1088/0004-637X/726/2/57},
  \href {https://ui.adsabs.harvard.edu/abs/2011ApJ...726...57C} {726, 57}

\bibitem[\protect\citeauthoryear{{De Rosa} et~al.,}{{De Rosa}
  et~al.}{2019}]{De_Rosa_2019}
{De Rosa} A.,  et~al., 2019, \mn@doi [\nar] {10.1016/j.newar.2020.101525},
  \href {https://ui.adsabs.harvard.edu/abs/2019NewAR..8601525D} {86, 101525}

\bibitem[\protect\citeauthoryear{{Di Matteo}, {Springel}  \& {Hernquist}}{{Di
  Matteo} et~al.}{2005}]{Di_Matteo_2005}
{Di Matteo} T.,  {Springel} V.,   {Hernquist} L.,  2005, \mn@doi [\nat]
  {10.1038/nature03335}, \href
  {https://ui.adsabs.harvard.edu/abs/2005Natur.433..604D} {433, 604}

\bibitem[\protect\citeauthoryear{{Ellison}, {Viswanathan}, {Patton},
  {Bottrell}, {McConnachie}, {Gwyn}  \& {Cuillandre}}{{Ellison}
  et~al.}{2019}]{Ellison_2019}
{Ellison} S.~L.,  {Viswanathan} A.,  {Patton} D.~R.,  {Bottrell} C.,
  {McConnachie} A.~W.,  {Gwyn} S.,   {Cuillandre} J.-C.,  2019, \mn@doi
  [\mnras] {10.1093/mnras/stz1431}, \href
  {https://ui.adsabs.harvard.edu/abs/2019MNRAS.487.2491E} {487, 2491}

\bibitem[\protect\citeauthoryear{{Kocevski} et~al.,}{{Kocevski}
  et~al.}{2017}]{Kocevski_2017}
{Kocevski} D.~D.,  et~al., 2017, \mn@doi [\apj] {10.3847/1538-4357/aa8566},
  \href {https://ui.adsabs.harvard.edu/abs/2017ApJ...846..112K} {846, 112}

\bibitem[\protect\citeauthoryear{{Koss}, {Mushotzky}, {Treister}, {Veilleux},
  {Vasudevan}  \& {Trippe}}{{Koss} et~al.}{2012}]{Koss_2012}
{Koss} M.,  {Mushotzky} R.,  {Treister} E.,  {Veilleux} S.,  {Vasudevan} R.,
  {Trippe} M.,  2012, \mn@doi [\apjl] {10.1088/2041-8205/746/2/L22}, \href
  {https://ui.adsabs.harvard.edu/abs/2012ApJ...746L..22K} {746, L22}

\bibitem[\protect\citeauthoryear{{Lemon}, {Auger}, {McMahon}  \&
  {Ostrovski}}{{Lemon} et~al.}{2018}]{Lemon_2018}
{Lemon} C.~A.,  {Auger} M.~W.,  {McMahon} R.~G.,   {Ostrovski} F.,  2018,
  \mn@doi [\mnras] {10.1093/mnras/sty911}, \href
  {https://ui.adsabs.harvard.edu/abs/2018MNRAS.479.5060L} {479, 5060}

\bibitem[\protect\citeauthoryear{{Lena}, {Panizo-Espinar}, {Jonker}, {Torres}
  \& {Heida}}{{Lena} et~al.}{2018}]{Lena_2018}
{Lena} D.,  {Panizo-Espinar} G.,  {Jonker} P.~G.,  {Torres} M.~A.~P.,   {Heida}
  M.,  2018, \mn@doi [\mnras] {10.1093/mnras/sty1147}, \href
  {https://ui.adsabs.harvard.edu/abs/2018MNRAS.478.1326L} {478, 1326}

\bibitem[\protect\citeauthoryear{{McKean}, {Berciano Alba}, {Volino}, {Tudose},
  {Garrett}, {Loenen}, {Paragi}  \& {Wucknitz}}{{McKean}
  et~al.}{2011}]{McKean_2011}
{McKean} J.~P.,  {Berciano Alba} A.,  {Volino} F.,  {Tudose} V.,  {Garrett}
  M.~A.,  {Loenen} A.~F.,  {Paragi} Z.,   {Wucknitz} O.,  2011, \mn@doi
  [\mnras] {10.1111/j.1745-3933.2011.01043.x}, \href
  {https://ui.adsabs.harvard.edu/abs/2011MNRAS.414L..11M} {414, L11}

\bibitem[\protect\citeauthoryear{{Millon} et~al.,}{{Millon}
  et~al.}{2020}]{Millon_2020}
{Millon} M.,  et~al., 2020, \mn@doi [\aap] {10.1051/0004-6361/202037740}, \href
  {https://ui.adsabs.harvard.edu/abs/2020A&A...640A.105M} {640, A105}

\bibitem[\protect\citeauthoryear{{Oguri}}{{Oguri}}{2010}]{Oguri_2010}
{Oguri} M.,  2010, {glafic: Software Package for Analyzing Gravitational
  Lensing} (\mn@eprint {ascl} {1010.012})

\bibitem[\protect\citeauthoryear{{Peng}, {Ho}, {Impey}  \& {Rix}}{{Peng}
  et~al.}{2010}]{Peng_2010}
{Peng} C.~Y.,  {Ho} L.~C.,  {Impey} C.~D.,   {Rix} H.-W.,  2010, \mn@doi [\aj]
  {10.1088/0004-6256/139/6/2097}, \href
  {https://ui.adsabs.harvard.edu/abs/2010AJ....139.2097P} {139, 2097}

\bibitem[\protect\citeauthoryear{{Planck Collaboration} et~al.,}{{Planck
  Collaboration} et~al.}{2016}]{planck2016}
{Planck Collaboration} et~al., 2016, \mn@doi [\aap]
  {10.1051/0004-6361/201525830}, \href
  {https://ui.adsabs.harvard.edu/abs/2016A&A...594A..13P} {594, A13}

\bibitem[\protect\citeauthoryear{{Raffai}, {Haiman}  \& {Frei}}{{Raffai}
  et~al.}{2016}]{Raffai_2015}
{Raffai} P.,  {Haiman} Z.,   {Frei} Z.,  2016, \mn@doi [\mnras]
  {10.1093/mnras/stv2371}, \href
  {https://ui.adsabs.harvard.edu/abs/2016MNRAS.455..484R} {455, 484}

\bibitem[\protect\citeauthoryear{{Rosas-Guevara}, {Bower}, {McAlpine}, {Bonoli}
   \& {Tissera}}{{Rosas-Guevara} et~al.}{2019}]{Rosas_Guevara_2018}
{Rosas-Guevara} Y.~M.,  {Bower} R.~G.,  {McAlpine} S.,  {Bonoli} S.,
  {Tissera} P.~B.,  2019, \mn@doi [\mnras] {10.1093/mnras/sty3251}, \href
  {https://ui.adsabs.harvard.edu/abs/2019MNRAS.483.2712R} {483, 2712}

\bibitem[\protect\citeauthoryear{{Rubinur}, {Das}  \& {Kharb}}{{Rubinur}
  et~al.}{2019}]{Rubinur_2019}
{Rubinur} K.,  {Das} M.,   {Kharb} P.,  2019, \mn@doi [\mnras]
  {10.1093/mnras/stz334}, \href
  {https://ui.adsabs.harvard.edu/abs/2019MNRAS.484.4933R} {484, 4933}

\bibitem[\protect\citeauthoryear{{Rusu}, {Berghea}, {Fassnacht}, {More},
  {Seman}, {Nelson}  \& {Chen}}{{Rusu} et~al.}{2019}]{Rusu_2019}
{Rusu} C.~E.,  {Berghea} C.~T.,  {Fassnacht} C.~D.,  {More} A.,  {Seman} E.,
  {Nelson} G.~J.,   {Chen} G. C.~F.,  2019, \mn@doi [\mnras]
  {10.1093/mnras/stz1142}, \href
  {https://ui.adsabs.harvard.edu/abs/2019MNRAS.486.4987R} {486, 4987}

\bibitem[\protect\citeauthoryear{{Sanders}, {Soifer}, {Elias}, {Madore},
  {Matthews}, {Neugebauer}  \& {Scoville}}{{Sanders}
  et~al.}{1988}]{Sanders_1988}
{Sanders} D.~B.,  {Soifer} B.~T.,  {Elias} J.~H.,  {Madore} B.~F.,  {Matthews}
  K.,  {Neugebauer} G.,   {Scoville} N.~Z.,  1988, \mn@doi [\apj]
  {10.1086/165983}, \href
  {https://ui.adsabs.harvard.edu/abs/1988ApJ...325...74S} {325, 74}

\bibitem[\protect\citeauthoryear{{Scott} \& {Kaviraj}}{{Scott} \&
  {Kaviraj}}{2014}]{Scott_2014}
{Scott} C.,  {Kaviraj} S.,  2014, \mn@doi [\mnras] {10.1093/mnras/stt2014},
  \href {https://ui.adsabs.harvard.edu/abs/2014MNRAS.437.2137S} {437, 2137}

\bibitem[\protect\citeauthoryear{{Shah} et~al.,}{{Shah}
  et~al.}{2020}]{Shah_2020}
{Shah} E.~A.,  et~al., 2020, \mn@doi [\apj] {10.3847/1538-4357/abbf59}, \href
  {https://ui.adsabs.harvard.edu/abs/2020ApJ...904..107S} {904, 107}

\bibitem[\protect\citeauthoryear{{Shajib} et~al.,}{{Shajib}
  et~al.}{2019}]{Shajib_2019}
{Shajib} A.~J.,  et~al., 2019, \mn@doi [\mnras] {10.1093/mnras/sty3397}, \href
  {https://ui.adsabs.harvard.edu/abs/2019MNRAS.483.5649S} {483, 5649}

\bibitem[\protect\citeauthoryear{{Sharma} et~al.,}{{Sharma}
  et~al.}{2021}]{Sharma_2021}
{Sharma} R.~S.,  et~al., 2021, arXiv e-prints, \href
  {https://ui.adsabs.harvard.edu/abs/2021arXiv210101729S} {p. arXiv:2101.01729}

\bibitem[\protect\citeauthoryear{{Shen} et~al.,}{{Shen}
  et~al.}{2021}]{Shen_2021}
{Shen} L.,  et~al., 2021, \mn@doi [\apj] {10.3847/1538-4357/abee75}, \href
  {https://ui.adsabs.harvard.edu/abs/2021ApJ...912...60S} {912, 60}

\bibitem[\protect\citeauthoryear{{Shields} \& {Bonning}}{{Shields} \&
  {Bonning}}{2008}]{Shields_2008}
{Shields} G.~A.,  {Bonning} E.~W.,  2008, \mn@doi [\apj] {10.1086/589427},
  \href {https://ui.adsabs.harvard.edu/abs/2008ApJ...682..758S} {682, 758}

\bibitem[\protect\citeauthoryear{{Silva}, {Marchesini}, {Silverman}, {Martis},
  {Iono}, {Espada}  \& {Skelton}}{{Silva} et~al.}{2021}]{Silva_2021}
{Silva} A.,  {Marchesini} D.,  {Silverman} J.~D.,  {Martis} N.,  {Iono} D.,
  {Espada} D.,   {Skelton} R.,  2021, \mn@doi [\apj]
  {10.3847/1538-4357/abdbb1}, \href
  {https://ui.adsabs.harvard.edu/abs/2021ApJ...909..124S} {909, 124}

\bibitem[\protect\citeauthoryear{{Solanes}, {Perea}, {Valent{\'\i}-Rojas}, {del
  Olmo}, {M{\'a}rquez}, {Ramos Almeida}  \& {Tous}}{{Solanes}
  et~al.}{2019}]{Solanes_2019}
{Solanes} J.~M.,  {Perea} J.~D.,  {Valent{\'\i}-Rojas} G.,  {del Olmo} A.,
  {M{\'a}rquez} I.,  {Ramos Almeida} C.,   {Tous} J.~L.,  2019, \mn@doi [\aap]
  {10.1051/0004-6361/201833767}, \href
  {https://ui.adsabs.harvard.edu/abs/2019A&A...624A..86S} {624, A86}

\bibitem[\protect\citeauthoryear{{Spilker} et~al.,}{{Spilker}
  et~al.}{2016}]{Spilker_2016}
{Spilker} J.~S.,  et~al., 2016, \mn@doi [\apj] {10.3847/0004-637X/826/2/112},
  \href {http://adsabs.harvard.edu/abs/2016ApJ...826..112S} {826, 112}

\bibitem[\protect\citeauthoryear{{Spingola} et~al.,}{{Spingola}
  et~al.}{2020}]{Spingola_2020}
{Spingola} C.,  et~al., 2020, \mn@doi [\mnras] {10.1093/mnras/staa1342}, \href
  {https://ui.adsabs.harvard.edu/abs/2020MNRAS.495.2387S} {495, 2387}

\bibitem[\protect\citeauthoryear{{Stacey} et~al.,}{{Stacey}
  et~al.}{2018}]{Stacey_2018}
{Stacey} H.~R.,  et~al., 2018, \mn@doi [\mnras] {10.1093/mnras/sty458}, \href
  {https://ui.adsabs.harvard.edu/abs/2018MNRAS.476.5075S} {476, 5075}

\bibitem[\protect\citeauthoryear{{Volonteri}, {Haardt}  \& {Madau}}{{Volonteri}
  et~al.}{2003}]{Volonteri_2003}
{Volonteri} M.,  {Haardt} F.,   {Madau} P.,  2003, \mn@doi [\apj]
  {10.1086/344675}, \href
  {https://ui.adsabs.harvard.edu/abs/2003ApJ...582..559V} {582, 559}

\bibitem[\protect\citeauthoryear{{Volonteri} et~al.,}{{Volonteri}
  et~al.}{2020}]{Volonteri_2020}
{Volonteri} M.,  et~al., 2020, \mn@doi [\mnras] {10.1093/mnras/staa2384}, \href
  {https://ui.adsabs.harvard.edu/abs/2020MNRAS.498.2219V} {498, 2219}

\bibitem[\protect\citeauthoryear{{Weston}, {McIntosh}, {Brodwin}, {Mann},
  {Cooper}, {McConnell}  \& {Nielsen}}{{Weston} et~al.}{2017}]{Weston_2016}
{Weston} M.~E.,  {McIntosh} D.~H.,  {Brodwin} M.,  {Mann} J.,  {Cooper} A.,
  {McConnell} A.,   {Nielsen} J.~L.,  2017, \mn@doi [\mnras]
  {10.1093/mnras/stw2620}, \href
  {https://ui.adsabs.harvard.edu/abs/2017MNRAS.464.3882W} {464, 3882}

\bibitem[\protect\citeauthoryear{{White} \& {Rees}}{{White} \&
  {Rees}}{1978}]{White_1978}
{White} S.~D.~M.,  {Rees} M.~J.,  1978, \mn@doi [\mnras]
  {10.1093/mnras/183.3.341}, \href
  {https://ui.adsabs.harvard.edu/abs/1978MNRAS.183..341W} {183, 341}

\makeatother
\end{thebibliography}




\bsp	
\label{lastpage}
\end{document}